\documentclass[fleqn]{NMD_of_RMF}
\usepackage{soul, color, xcolor}
\soulregister{\cite}7 % 注册\cite命令
\soulregister{\citep}7 % 注册\citep命令
\soulregister{\citet}7 % 注册\citet命令
\soulregister{\ref}7 % 注册\ref命令
\soulregister{\pageref}7 % 注册\pageref命令
\usepackage{epstopdf}
\begin{document}

\ensubject{subject}

%%%%%%%%%%%%%%%%%%%%%%%%%%%%%%%%%%%%%%%%%%%%%%%%%%%%%%%
%%% Authors do not modify the information below
%%% ????????????????
%%% ??????????, ????????????{}, ???????????????????
%Letter to the Editor??Article%??????
\ArticleType{Article}%??Article
\SpecialTopic{SPECIAL TOPIC: }%???????
\Year{2023}
\Month{January}
%\Vol{63}
%\No{1}
%\DOI{??}
%\ArtNo{000000}
%\ReceiveDate{January 11, 2020}
%\AcceptDate{April 6, 2020}
%\OnlineDate{January 1, 2016}
%%%%%%%%%%%%%%%%%%%%%%%%%%%%%%%%%%%%%%%%%%%%%%%%%%%%%%%

%%% title: ????
%%%   \title{title}{title for citation}
\title{New extended method for $\psi^{\prime}$ scaling function of inclusive electron scattering}

%%% Corresponding author: ???????
%%%   \author[number]{Full name}{{email@xxx.com}}
%%% General author: ???????
%%%   \author[number]{Full name}{}
\author[1]{Lei Wang }{}%
\author[2]{Qinglin Niu}{}%
\author[3]{Jinjuan Zhang}{}% 
\author[2,4,5]{Jian Liu}{liujian@upc.edu.cn}%
\author[1,6]{Zhongzhou Ren }{zren@tongji.edu.cn}%

%%% Author information for page head. ??????????
%%% ??????????????, ??????????author???
\AuthorMark{Lei Wang}%\authorcr????????

%%% Authors for citation. ?????????????????
%%% ??????????????, ??????????author???
\AuthorCitation{L.~Wang, Q.~Niu, J.~Zhang, J.~Liu and Z.~Ren, New extended method for $\psi^{\prime}$ scaling function of inclusive electron scattering}

%%% Address. ???
%%%   \address[number]{Address, City {\rm Postcode}, Country}
\address[1]{School of Physics Science and Engineering, Tongji University, Shanghai 200092, China}
\address[2]{College of Science,  China University of Petroleum (East China), Qingdao 266580, China}
\address[3]{College of Electronic and Information Engineering, Shandong University of Science and Technology, Qingdao 266590, China}
\address[4]{Laboratory of High Precision Nuclear Spectroscopy, Insititute of Modern Physics, Chinese Academy of Sciences, Lanzhou 730000, China}
\address[5]{Guangxi Key Laboratory of Nuclear Physics and Nuclear Technology, Guangxi Normal University, Guilin 541004, China}
\address[6]{Key Laboratory of Advanced Micro-Structure Materials, Ministry of Education, Shanghai 200092, China}

%\contributions{}%????????

%%% Abstract. ??
\abstract{Scaling analyses have been successfully applied to study the inclusive electron scattering $(\textit{e}, \textit{e}^{\prime})$ over the past few decades. In this paper, we utilize the $\psi^{\prime}$ scaling function in momentum space to analyze the $(\textit{e}, \textit{e}^{\prime})$ cross sections, where the nucleon momentum distributions are derived from self-consistent mean-field calculations. By further introducing the energy and momentum conservation in the scaling analysis, an improved  $\psi^{\prime}$ scaling function is proposed to investigate the high-momentum part of the momentum distributions. Using the proposed scaling function, we systematically explore the effects of the nucleon-nucleon short-range correlation ($\textit{NN}$-SRC) on the 	$(\textit{e}, \textit{e}^{\prime})$ cross sections. From the experimental $(\textit{e}, \textit{e}^{\prime})$ data, the \textit{NN}-SRC strength is further extracted within the framework of the improved $\psi^{\prime}$ scaling function. The studies in this paper offer a new method to investigate the nucleon momentum distributions and the $\textit{NN}$-SRC effects in nuclei.}%??
%Scaling analyses have been successfully applied in reproducing the inclusive electron scattering $(\textit{e}, \textit{e}^{\prime})$ data during the past decades. In this paper, utilizing the coherent density fluctuation model (CDFM), the $\psi^{\prime}$ scaling function is investigated by the nucleon momentum distributions $n(k)$ from the self-consistent mean-field calculations. Further considering the momentum and energy conservation in the $(\textit{e}, \textit{e}^{\prime})$ process, we present an improved  $\psi^{\prime}$ scaling function to investigate the high-momentum part of the $n(k)$. With this improved method, the sensitivity of nucleon-nucleon short-range correlation ($\textit{NN}$-SRC) effects on $(\textit{e}, \textit{e}^{\prime})$ cross sections is analyzed systematically. Besides, the $\textit{NN}$-SRC strength is extracted from the experimental data based on the improved $\psi^{\prime}$ scaling function. The studies in this paper offer a new method to investigate the nucleon momentum distributions and the $\textit{NN}$-SRC influences.
%%% Keywords. ?????


\maketitle

%\tableofcontents%?????

%%%%%%%%%%%%%%%%%%%%%%%%%%%%%%%%%%%%%%%%%%%%%%%%%%%%%%%
%%% The main text. ???????
%???????????????????\cref{fig1}
%\twocolumn\onecolumn
%%%%%%%%%%%%%%%%%%%%%%%%%%%%%%%%%%%%%%%%%%%%%%%%%%%%%%%
\begin{multicols}{2}
\section{Introduction}\label{section1}
Quasielastic (QE) electron scattering, which has aroused considerable attention in the last decades, provides a crucial window to study nuclear structure beyond the mean-field picture \cite{Benhar2008Inclusive}. From the QE scattering, the information of the nucleon momentum distribution $n(k)$ can be extracted from the shape and width of the QE peak \cite{CLAS2019Modified,Duer2018probing,Schmidt2020Probing,AntonovSuperscaling2007,IVANOV2020nucleon,antonov1988nucleon}.
%the information of nucleon momentum distribution $n(k)$ can be extracted from observable cross sections \cite{CLAS2019Modified,Duer2018probing,Schmidt2020Probing,AntonovSuperscaling2007,IVANOV2020nucleon,antonov1988nucleon}. 
 The high-momentum components of the momentum distributions offer insight into the nucleon-nucleon short-range ($\textit{NN}$-SRC) effects \cite{Hen2014Momentum}, which play a vital role in comprehending the characteristics of matter at high densities, such as relativistic heavy-ion scattering and neutron stars \cite{degli2015medium,Hen2015,Akimov2017Observation,Hen2017Nucleon,LI2018Nucleon,xue2016neutron,zhang2016measurements,Wang2017Probing,Wan2020Finite,wei2022effects,liu2022density,Yuan2023myz}.

Significant advancements were witnessed in the field of quasielastic electron scattering experiments at Jefferson
Laboratory \cite{Duer2019Direct,Egiyan2006Measurement}. There are two groups of quasielastic scattering experiments: exclusive electron scattering ($\textit{e}, \textit{e}^{\prime}\textit{p}\textit{n}$) and inclusive electron scattering $(\textit{e}, \textit{e}^{\prime})$. From the ($\textit{e}, \textit{e}^{\prime}\textit{p}\textit{n}$) measurement, it is indicated that the neutron-proton ($\textit{n}$-$\textit{p}$) pairs are approximately 20 times as many as neutron-neutron ($\textit{n}$-$\textit{n}$) pairs and proton-proton ($\textit{p}$-$\textit{p}$) pairs in $^{12}\rm{C}$ \cite{Subedi2008Probing}, which can be attributed to the effects of tensor force in the calculations of light nuclei with realistic nucleon-nucleon potential \cite{Schiavilla2007Tensor}. From the $(\textit{e}, \textit{e}^{\prime})$ experiments \cite{Murphy2019Measurement,Dai2018First,Dai2019First,Ye2018Search}, the $\textit{NN}$-SRC strengths have been extracted from the ratios of cross sections of heavy nuclei to those of the deuteron, at the squared four-momentum transfer $Q^2>1.4$~$\rm{GeV}^2$ \cite{Frankfurt1993Evidence,Fomin2012New}. Based on the High Intensity heavy-ion Accelerator Facility (HIAF), the Electron-Ion Collider in China (EicC) project will be constructed with a new electron ring in the near future \cite{anderle2021electron}, which is essential  for studies of  high-energy electron scattering, as well as the nuclear and nucleon structures.

%It will be beneficial for studies of the nuclear and nucleon structure \cite{anderle2021electron}, as well as detector and accelerator technology.% In addition, the new accelerator of quasi-elastic electron scattering experiments will be constructed around the world \cite{anderle2021electron}.%Based on the High Intensity heavy-ion Accelerator Facility (HIAF), the Electron-Ion Collider in China (EicC) project will be constructed with a new electron ring in the near future, which will be beneficial for studies of the nuclear and nucleon structure \cite{anderle2021electron}, as well as detector and accelerator technology.

In view of these experiments, plenty of theoretical researches on quasielastic electron scattering are developed accordingly \cite{Benhar2010Electroweak,Ciofi1996Realistic,Meucci2013Elastic,Niu2022Effects,Barbaro2019mean}. As a powerful tool to analyze the $(\textit{e}, \textit{e}^{\prime})$ experimental data, the scaling method is derived from the plane-wave impulse approximation (PWIA) \cite{day1990scaling,Ciofi1991y-scaling,tDonnelly1999Superscaling}.
%One powerful tool is the scaling method \cite{day1990scaling,Ciofi1991y-scaling,tDonnelly1999Superscaling}, which is derived from the plane-wave impulse approximation (PWIA). 
% lei4.18 Within the scaling scheme, the theoretical inclusive cross sections can be expressed as a product of two parts: the single-nucleon cross sections $\sigma_{eN}$ and a scaling function. $\sigma_{eN}$ represents the scattering cross sections of an electron by a off-shell nucleon. The scaling function contains the information about the dynamical properties of the nuclear ground state. 
% and the dynamics of the interacting system. The information about the dynamical properties of the nuclear ground state is included in the scaling function.
%in which the information about the dynamical properties of the nuclear ground state is included in the scaling function. 
Within the PWIA scheme, the theoretical inclusive cross sections can be expressed as a product of two parts: the single-nucleon cross sections $\sigma_{eN}$ and the scaling function. $\sigma_{eN}$ represents the scattering cross sections of an electron by an off-shell nucleon. The scaling function contains the information about dynamical properties of the nuclear ground state. There are different scaling variables, such as $x$ scaling \cite{Kendall1991Deep}, $y$ scaling \cite{Ciofi1991y-scaling}, and $\psi^{\prime}$ scaling \cite{Donnelly1999Superscaling}. 
%There are different scaling variables, such as $x$ scaling, $y$ scaling and $\psi^{\prime}$ scaling 
As a superscaling variable, the $\psi^{\prime}$ scaling  is dimensionless and independent of the momentum transfer $q$ as well as the mass number $A$. It is noted that the $\psi^{\prime}$ scaling function $f(\psi^{\prime})$ is successful in investigating the quasielastic electron- and neutrino-nucleus scattering \cite{Megias2015Meson,Liang2022Nucleon}.

%in reproducing the $(\textit{e}, \textit{e}^{\prime})$ cross sections for both quasi-elastic and $\Delta$ regions. Recently, $\psi^{\prime}$ scaling is widely applied to the field of lepton scattering \cite{Megias2015Meson}.  

The $\psi^{\prime}$ scaling function was first performed based on the relativistic Fermi gas (RFG) model \cite{Alberico1988Scaling}. 
However, compared with the experimental data, the RFG model fails to provide accurate descriptions of the $(\textit{e}, \textit{e}^{\prime})$ cross sections, and the inclusion of detailed calculations of nuclear structure is required in the scaling function \cite{Antonov2006Superscaling}. As a natural extension of the RFG model to realistic nuclear systems, the coherent density fluctuation model (CDFM) introduces the nuclear density distributions on the basis of the generator coordinate method \cite{Antonov2004Superscaling}. Within the CDFM, the scaling function can be connected with the microscopic nuclear structure models, and the nucleon distributions inside nuclei can be reflected from the $(\textit{e}, \textit{e}^{\prime})$ cross sections \cite{Antonov2005Superscaling,Antonov2006Scaling}. 

In previous studies, the scaling function has been systematically investigated from the nuclear density distribution $\rho(r)$ with CDFM in coordinate space ($r$-space) \cite{Antonov2004Superscaling,Wang2021Global}. It is realized that the nucleon momentum distribution $n(k)$ is more sensitive to the $\textit{NN}$-SRC effects than the density distribution $\rho(r)$ \cite{degli2015medium,sun2016green}. Therefore, in order to better analyze the $\textit{NN}$-SRC effects, it is necessary to provide a scaling function based on the momentum distribution $n(k)$. For this purpose, this paper utilizes the $\psi^{\prime}$ scaling function in momentum space ($k$-space) $f_{n(k)}^{\rm{Q E}}\left(\psi^{\prime}\right)$ to investigate the $(\textit{e}, \textit{e}^{\prime})$ process. To analyze the effects of different momentum distributions on the $(\textit{e}, \textit{e}^{\prime})$ cross sections, an improved $\psi^{\prime}$ scaling function $\mathscr{F}_{n(k)}^{\rm{QE}}\left(\psi^{\prime}\right)$ is presented by introducing the momentum and energy conservation. 
The main work and contribution of this paper are as follows. Firstly, the nucleon momentum distributions $n(k)$ are investigated by combining the axially deformed Hartree-Fock-Bogoliubov (HFB) model with the light-front dynamics (LFD) method.  Secondly, with the momentum distributions $n(k)$, the scaling function in $k$-space $f_{n(k)}^{\rm{Q E}}\left(\psi^{\prime}\right)$ is investigated to calculate the corresponding $(\textit{e}, \textit{e}^{\prime})$ cross sections. Then, on the basis of $f_{n(k)}^{\rm{Q E}}\left(\psi^{\prime}\right)$, the improved scaling function $\mathscr{F}_{n(k)}^{\rm{QE}}\left(\psi^{\prime}\right)$  is presented by including the momentum and energy conservation of $(\textit{e}, \textit{e}^{\prime})$ process. 
%By this means, $\textit{NN}$-SRC effects can be reflected in the scaling function. 
%The comparison between $f_{n(k)}^{\rm{Q E}}\left(\psi^{\prime}\right)$ and $\mathscr{F}_{n(k)}^{\rm{QE}}\left(\psi^{\prime}\right)$ is also provided. 
Finally, the effects of high-momentum components of nucleon momentum distributions on the $(\textit{e}, \textit{e}^{\prime})$ cross sections are analyzed and the correlation strength is extracted theoretically.

This paper is organized as follows: In Sec.~\ref{sec:2}, the theoretical frameworks of the deformed HFB model and improved CDFM method are provided. In Sec.~\ref{sec:3}, the numerical results and discussions are presented. A summary is given in Sec.~\ref{sec:4}.

\section{Theoretical framework}\label{sec:2}
In this section, we provide the $\psi^{\prime}$ scaling function in momentum space. Considering the energy and momentum conservation of inclusive electron scattering, an improved $\psi^{\prime}$ scaling function is constructed in this part.  
\subsection{$\psi^{\prime}$ scaling function in $k$-space}
On the basis of PWIA, the inclusive cross sections can be organized into the product of the single-nucleon cross section and a scaling function, which is generally a function of electron momentum transfer $q$ and energy transfer $\omega$. In the limit of large $q$, the scaling function depends on only a single variable named the scaling variable, such as $\psi^{\prime}$ variable. In the QE region, the $\psi^{\prime}$ scaling is introduced in Ref. \cite{Donnelly1999Superscaling,Maieron2009Superscaling}
\begin{equation}\label{1}
	\psi^{\prime} \equiv \frac{1}{\sqrt{\xi_F}} \frac{\lambda^{\prime}-\tau^{\prime}}{\sqrt{\left(1+\lambda^{\prime}\right) \tau^{\prime}+\kappa \sqrt{\tau^{\prime}\left(\tau^{\prime}+1\right)}}},
\end{equation}
where 
	\begin{align}
		\xi_F & \equiv \sqrt{1+\eta_F^2}-1, \quad \eta_F \equiv \frac{k_F}{m_N},\label{kf}\\
		\lambda^{\prime} &=\lambda-\frac{E_{\text {shift }}}{2 m_N}, \quad \lambda=\frac{\omega}{2 m_N},\label{22}\\
		\tau^{\prime}&=\kappa^2-\lambda^{\prime 2},\quad \kappa=\frac{q}{2 m_N}.
	\end{align}
 In natural unit system, the above quantities $\xi_F$, $\eta_F$, $\lambda$, $\kappa$, and $\tau^{\prime}$ are utilized  to provide a convenient dimensionless scale in quasielastic electron scattering. $m_N$ is the nucleon mass. $E_{\text{shift}}$ of Eq.~(\ref{22}) is the empirical parameter to adjust the position of QE peak \cite{Maieron2002Extended}. $k_F$ in Eq.~(\ref{kf}) is the nuclear Fermi momentum defined as 
\begin{equation}
	k_F=\frac{16 \pi}{3 A} \int_0^{\infty} d k~n\left(k\right) k^3,
\end{equation}
where $k$ is the nucleon momentum.

In momentum space ($k$-space), the scaling function can be derived from the relativistic Fermi gas (RFG) model analytically as \cite{Antonov2006Superscaling}
\begin{equation}\label{3}
	\begin{aligned}
		f_{\mathrm{RFG}}\left(k, \psi^{\prime}\right)=& \frac{3}{4}\left[1-\left(\frac{k_F\left|\psi^{\prime}\right|}{k}\right)^2\right]  \times\left\{1+\left(\frac{m_N}{k}\right)^2\left(\frac{k_F\left|\psi^{\prime}\right|}{k}\right)^2\right.\\
		&\left.\times\left[2+\left(\frac{k}{m_N}\right)^2-2 \sqrt{1+\left(\frac{k}{m_N}\right)^2}~\right]\right\}.
	\end{aligned}
\end{equation}
%As an extension of the relativistic Fermi gas (RFG) model, CDFM is proposed to study the scaling function, where the nuclear structure information is contained. Based on the momentum distribution $n(k)$, the QE scaling function can be expressed as follows \cite{Antonov2006Superscaling}
%The CDFM is proposed as an extension of the RFG model for investigating the scaling function, which contains the nuclear structure information. Based on the $n(k)$, the QE scaling function can be expressed as follows \cite{Antonov2006Superscaling}
As a development of RFG, the CDFM scaling function is proposed to include the nuclear structure information. With the CDFM, the QE scaling function in $k$-space can be expressed as follows \cite{Antonov2006Superscaling}
\begin{equation}\label{2}
	f_{n(k)}^{\rm{Q E}}\left(\psi^{\prime}\right)=\int_{k_F\left|\psi^{\prime}\right|}^{\infty} d k\left|G\left(k\right)\right|^2 f_{\mathrm{RFG}}\left(k, \psi^{\prime}\right),
\end{equation}
%where 
%
%$m_N$ is the nucleon mass, $k$ is the nucleon momentum, and $k_F$ is Fermi momentum and calculated by
%\begin{equation}
%	k_F=\frac{16 \pi}{3 A} \int_0^{\infty} d k~n\left(k\right) k^3.
%\end{equation}
where $\left|G\left(k\right)\right|^2$ is the weight function and closely related to momentum distribution $n(k)$
\begin{equation}\label{4}
	\left|G\left(k\right)\right|^2=-\left.\frac{1}{n_0\left(k\right)} \frac{d n(k)}{d k}\right., \quad n_0\left(k\right)=\frac{3 A}{4 \pi k^3} .
\end{equation}
$n_0(k)$ represents the density distribution that contains all $A$ nucleons distributed homogeneously in a sphere of $k$-space.

In this paper, we extend our scaling analysis to the $\Delta$ region. The scaling variable in the $\Delta$ region is defined as \cite{Amaro2005Using}
\begin{equation}
\begin{aligned}
	\psi_{\Delta}^{\prime} \equiv & {\left[\frac{1}{\xi_F}\left(\kappa \sqrt{\rho_{\Delta}^{\prime 2}+1 / \tau^{\prime}}-\lambda^{\prime} \rho_{\Delta}^{\prime}-1\right)\right]^{1 / 2} } \\
	& \times \begin{cases}+1, & \lambda^{\prime} \geqslant \lambda_{\Delta}^{\prime 0} \\
		-1, & \lambda^{\prime} \leqslant \lambda_{\Delta}^{\prime 0}\end{cases}~~,
\end{aligned}
\end{equation}
where the dimensionless variables are
\begin{align}
	\lambda_{\Delta}^{\prime 0} & =\lambda_{\Delta}^0-\frac{E_{\text {shift }}}{2 m_N}, \quad \lambda_{\Delta}^0=\frac{1}{2}\left(\sqrt{\mu_{\Delta}^2+4 \kappa^2}-1\right), \\
	\mu_{\Delta} & =\frac{m_{\Delta}(1232 ~\text{MeV})}{m_N}, \quad
	\beta_{\Delta}  =\frac{1}{4}\left(\mu_{\Delta}^2-1\right),\\ \rho_{\Delta}& =1+\frac{\beta_{\Delta}}{\tau}, \quad \rho_{\Delta}^{\prime}=1+\frac{\beta_{\Delta}}{\tau^{\prime}}.
\end{align}
With CDFM, the $\psi^{\prime}$ scaling function in the $\Delta$ region can be written as 
%The $\Delta$-scaling function can be expressed as
\begin{equation}
	f^{\Delta}\left(\psi_{\Delta}^{\prime}\right)=\int_0^{\infty} d k\left|G\left(k\right)\right|^2 f_{\mathrm{RFG}}^{\Delta}\left[\psi_{\Delta}^{\prime}\left(k\right)\right],
\end{equation}\label{5}
where $f_{\mathrm{RFG}}^{\Delta}$ represents the analytical RFG scaling function in the $\Delta$ region \cite{Antonov2005Superscaling}
\begin{equation}\label{6}
	f_{\mathrm{RFG}}^{\Delta}\left(\psi_{\Delta}^{\prime}\right)=\frac{3}{4}\left(1-\psi_{\Delta}^{\prime 2}\right) \Theta\left(1-\psi_{\Delta}^{\prime 2}\right) .
\end{equation}
 $\Theta\:(x)$ is the Heaviside step function. The function $\Theta\:(x)$ equals  zero for $x<0$  and  one for $x>0$.

%%%%%%%%%%%%%%%%%%%%%%%%%%%%%%%%%%%%%%%%%%%%%%%%%%%%%%%%%%%%%%%	

%%%%%%%%%%%%%%%%%%%%%%%%%%%%%%%%%%%%%%%%%%%%%%%%%%%%%%%%%%%%%%%	
\subsection{Improved $\psi^{\prime}$ scaling function in $k$-space}
%	由于对全空间积分，对动量分布的变化不明显
The contributions of nucleons with different momentum to the scaling function are reflected in Eq.~(\ref{2}). We further introduce the constrained condition of nucleon momentum by considering the conservation of the energy and momentum in $(\textit{e}, \textit{e}^{\prime})$ process \cite{Ciofi1991y-scaling}
%The integration limit in Eq.~(\ref{2}) is the approximate value $k_F\left|\psi^{\prime}\right|$, and the scaling function $f_{n(k)}^{\rm{Q E}}\left(\psi^{\prime}\right)$ cannot reflect the $\textit{NN}$-SRC influences accurately. This is due to the absence of energy and momentum conservation in Eq.~(\ref{2}) \cite{Ciofi1991y-scaling}
\begin{equation}\label{7}
	\begin{aligned}
		\omega+M_A^0= \sqrt{m_N^2+\left(\textbf{q} + \textbf{k}\right)^2} +\sqrt{\left(M_{A-1}^0+E_{A-1}^{f *}\right)^2+\textbf{k}^2},
	\end{aligned}
\end{equation}
where $M_A^0$ and $M_{A-1}^0$ represent the masses of the target and the recoil nuclei, respectively. $\textbf{q}$ and $\textbf{k}$ are the vector form of the momentum transfer and nucleon momentum. $E_{A-1}^{f *}$ denotes the excitation energy of the final $A-1$ nucleon system. Eq.~(\ref{7}) indicates that only nucleons satisfying the energy and momentum conservation can participate in the ($\textit{e},\textit{e}^{\prime}$) process at certain kinematic conditions.

%nucleons cannot fully participate in the ($\textit{e},\textit{e}^{\prime}$) process at certain kinematics.	
%The upper and lower limits of the nucleon momentum are fixed by Eq.~(\ref{7})

In Eq.~(\ref{7}), the angle $\alpha$ between $\textbf{q}$ and $\textbf{k}$ is $0-180^\circ$. Substituting $\alpha=0^\circ$ and $\alpha=180^\circ$ into Eq.~(\ref{7}), one can obtain the lower and upper limits of the nucleon momentum taking part in the $(\textit{e}, \textit{e}^{\prime})$ process 
%For $\alpha=0$ and $\alpha=180^\circ$, one can obtain the upper and lower limits of the nucleon momentum taking part in the $(\textit{e}, \textit{e}^{\prime})$ process 
%$\cos\alpha=(\textbf{q}\cdot\textbf{k})/(q\cdot k)=1~\text{or}-1~$, and the upper and lower limits of the nucleon momentum participating in the $(\textit{e}, \textit{e}^{\prime})$ process are determined as
\begin{subequations}\label{8}
	\begin{align}
		k_{\min }=~&\frac{1}{2 W^2}\bigg\{\left(M_A^0+\omega\right) \sqrt{W^2-\left(M_{A-1}^0+m_N\right)^2} \\ \nonumber&\times \sqrt{W^2-\left(M_{A-1}^0-m_N\right)^2}
		-q\left[W^2+\left(M_{A-1}^0\right)^2-m_N^2\right]\bigg\},\\ \nonumber\\
		k_{\max }=~&\frac{1}{2 W^2}\bigg\{\left(M_A^0+\omega\right) \sqrt{W^2-\left(M_{A-1}^0+m_N\right)^2} \\ \nonumber&\times \sqrt{W^2-\left(M_{A-1}^0-m_N\right)^2}
		+q\left[W^2+\left(M_{A-1}^0\right)^2-m_N^2\right]\bigg\},
	\end{align}
\end{subequations}
where $W=\sqrt{\left(M_A^0+\omega\right)^2-q^2}$.

Taking into account the conservation condition in Eq. (\ref{8}), the limit of the $\psi^{\prime}$ scaling function in the QE region of Eq. (\ref{2}) can be modified to the following expression
\begin{equation}\label{9}
	\mathscr{F}_{n(k)}^{\rm{QE}}\left(\psi^{\prime}\right)=\int_{k_{\min }}^{k_{\max }} dk\left|G\left(k\right)\right|^2 f_{\mathrm{RFG}}\left(k, \psi^{\prime}\right).
\end{equation}
Combining Eqs.~(\ref{7})~-~(\ref{9}), an improved QE scaling function in $k$-space $\mathscr{F}_{n(k)}^{\rm{QE}}\left(\psi^{\prime}\right)$ is proposed. Due to low-momentum nucleons being unable to contribute to the ($\textit{e},\textit{e}^{\prime}$) cross section, the improved scaling function $\mathscr{F}_{n(k)}^{\rm{QE}}\left(\psi^{\prime}\right)$ is sensitive to high-momentum nucleons, which can be utilized to investigate \textit{NN}-SRC effects.

%%%%%%%%%%%%%%%%%%%%%%%%%%%%%%%%%%%%%%%%%%%%%%%%%%%%%%%%%%%%%%%		
%%%%%%%%%%%%%%%%%%%%%%%%%%%%%%%%%%%%%%%%%%%%%%%%%%%%%%%%%%%%%%%	
\subsection{Inclusive cross section from scaling analyses}
Within the framework of scaling method, the double differential cross sections of inclusive electron scattering can be derived as 
\begin{equation}\label{10}
	\begin{aligned}
		\frac{d^2 \sigma}{d \Omega d \omega}=& \sigma_\text{M}\left[\left(\frac{Q^2}{q^2}\right)^2 R_L(q, \omega)\right.\left.+\left(\frac{1}{2}\left|\frac{Q^2}{q^2}\right|+\tan ^2 \frac{\theta}{2}\right) R_T(q, \omega)\right],
	\end{aligned}
\end{equation}
where the four-momentum transfer $Q^2=\omega^2-q^2$. $\sigma_\text{M}$ is the Mott cross section, which provides description of electron scattering off a positively point charge \cite{mott1929scattering}. 
%As longitudinal and transverse response functions, $R_L$ and $R_T$ contain the information about the nuclear electromagnetic charge distribution and current density distribution, which can be written as the following formalism in the QE and $\Delta$ regions 
In the QE region, the longitudinal and transverse response functions $R_L$ and $R_T$ can be expressed from the scaling function \cite{AMARO1999Relativistic}
\begin{subequations}\label{11}
	\begin{align}
		R_L^{\mathrm{QE}}(\kappa, \lambda)=&  \frac{\mathcal{N} \xi_F}{ m_N \eta_F^3 \kappa}\frac{\kappa^2}{\tau}\left[(1+\tau) W_2(\tau)-W_1(\tau)\right] \times f^{\mathrm{QE}}\left(\psi^{\prime}\right), \\
		R_T^{\mathrm{QE}}(\kappa, \lambda)=& \frac{\mathcal{N} \xi_F}{ m_N \eta_F^3 \kappa}2 W_1(\tau) \times f^{\mathrm{QE}}\left(\psi^{\prime}\right). \\ \nonumber
	\end{align}
\end{subequations}
where $\mathcal{N}$ is the proton number $Z$ or neutron number $N$. $W_1(\tau)$ and $W_2(\tau)$ represent the relativistic structure functions in the QE region, which can be deduced from the Sachs form factors $G_E(\tau)$ and $G_M(\tau)$ \cite{Antonov2006Superscaling}.

In the $\Delta$ region, the response functions are written as
\begin{subequations}\label{12}
	\begin{align}
		R_L^{\Delta}(\kappa, \lambda)=& \frac{\mathcal{N} \xi_F}{2 m_N \eta_F^3 \kappa} \frac{\kappa^2}{\tau}\bigg[\left(1+\tau \rho^2\right) w_2(\tau)-w_1(\tau)\\\nonumber
		&+w_2(\tau) D(\kappa, \lambda)\bigg] \times f^{\Delta}\left(\psi_{\Delta}^{\prime}\right), \\ 
		R_T^{\Delta}(\kappa, \lambda)=& \frac{\mathcal{N} \xi_F}{2 m_N \eta_F^3 \kappa}\left[2 w_1(\tau)+w_2(\tau) D(\kappa, \lambda)\right] \times f^{\Delta}\left(\psi_{\Delta}^{\prime}\right),
	\end{align}
\end{subequations}	
where 
\begin{equation}\label{21}
	\begin{aligned}
		D(\kappa, \lambda) \equiv & \frac{\tau}{\kappa^2}\bigg[(\lambda \rho+1)^2+(\lambda \rho+1)\left(1+\psi_{\Delta}^{\prime 2}\right) \xi_F \\
		& +\frac{1}{3}\left(1+\psi_{\Delta}^{\prime 2}+\psi_{\Delta}^{\prime 4}\right) \xi_F^2\bigg]-\left(1+\tau \rho^2\right),
	\end{aligned}
\end{equation}
%\begin{equation}
%\text{with} \quad \rho  =  1+\frac{1}{4 \tau}\left(\mu_{\Delta}^{2}-1\right). \nonumber
%\end{equation}
with $\rho=1+\frac{1}{4\tau}\left(\mu_{\Delta}^{2}-1\right)$. In the above equations, $w_1(\tau)$ and $w_2(\tau)$ are the structure functions in the $\Delta$ region, which are also associated with the Sachs form factors $G_E(\tau)$ and $G_M(\tau)$ \cite{Amaro2005Using}. Combining Eqs.~(\ref{10})~-~(\ref{21}), one can obtain the total inclusive cross sections containing the contributions of the QE and $\Delta$ regions.

%and $\mathcal{N}=N~\text{or}~Z$. $W_1(\tau)$ and $W_2(\tau)$ represent the relativistic structure functions in the QE region. $w_1(\tau)$ and $w_2(\tau)$ are the structure functions in the $\Delta$ region. These functions are associated with the Sachs form factors $G_E(\tau)$ and $G_M(\tau)$ \cite{Amaro2005Using,Antonov2006Superscaling}. Combining the Eqs.~(\ref{10})~-~(\ref{21}), one can obtain the total inclusive cross sections containing the contributions of the QE and $\Delta$ regions.

%%%%%%%%%%%%%%%%%%%%%%%%%%%%%%%%%%%%%%%%%%%%%%%%%%%%%%%%%%%%%%%	
~\\
\subsection{Nucleon momentum distribution}
In this part, nucleon momentum distributions (NMDs) are constructed by taking into account the contributions of single-particle motion and \textit{NN}-SRC effects
\begin{equation}\label{14}
	n(k)=n_{\mathrm{MF}}(k)+n_{\mathrm{corr}}(k),
\end{equation}
where $n_{\mathrm{MF}}(k)$ is the mean-field part ($k<k_F$) and $n_{\mathrm{corr}}(k)$ embodies \textit{NN}-SRC effects on high-momentum components ($k>k_F$). 

The Hartree-Fock-Bogoliubov (HFB) model, which has been extensively applied to study nuclear properties, is used to calculate the mean-field part $n_{\mathrm{MF}}(k)$ of NMDs \cite{Hern2021Nuclear,meng2020ground,Liu2022Exploring}. Starting from the energy density functional $\mathcal{H}(r)$, one can derive the HFB equation from the variation of total energy with respect to single-particle wave functions $\Phi_\alpha(\mathbf{r})$ \cite{wang2020charge,ge2020effects,Zhong2021yhm}. Solving the HFB equation iteratively, the nucleon wave functions in $r$-space $\Phi_\alpha(\boldsymbol{r}, \sigma)$ can be obtained. Through Fourier transformation, it is possible to derive the single-particle wave functions in $k$-space
\begin{equation}\label{23}
	\tilde{\Phi}_\alpha(\boldsymbol{k}, \sigma) = \frac{1}{(2 \pi)^{3 / 2}} \int d \boldsymbol{r}~ \mathrm{e}^{-i \boldsymbol{k} \cdot \boldsymbol{r}} \Phi_\alpha(\boldsymbol{r}, \sigma).
\end{equation}
With single-particle wave functions in $k$-space, one can construct the mean-field part of NMDs  \cite{stoitsov2007variation,MOYADEGUERRA1991Momentum}
\begin{equation}\label{13}
	n_{\mathrm{MF}}(\boldsymbol{k})=\sum_\alpha\left|\tilde{\Phi}_\alpha(\boldsymbol{k}, \sigma)\right|^2,
\end{equation}
which provides a good description for NMDs in the region $k<1.5~\mathrm{fm^{-1}}$.
The HFB code used in this paper allows for constrained axially symmetric deformations, in which the solution of the HFB equation is obtained by expanding the quasiparticle function in a complete set of harmonic oscillator (HO) basis wave functions.

Due to the lack of \textit{NN}-SRC effects, the mean-field model cannot reproduce the high-momentum component of NMDs. To import the effects of the
nucleon-nucleon correlations into the mean-field parts of NMDs, the light-front dynamics (LFD) method is introduced to calculate $n_{\mathrm{corr}}(k)$ by empirically rescaling the high-momentum components of NMDs of the deuteron  based on the natural orbital representation. In the LFD method, the the NMDs of the deuteron can be decomposed into six components $n_{1-6}(k)$ deduced from six LFD wave functions \cite{Antonov2002Nucleon}. In the region $k>1.5~\mathrm{fm^{-1}}$, the most crucial contributions to NMDs come from the components $n_2(k)$ and $n_5(k)$. Therefore, within the framework of LFD, the correlation parts of NMDs in nuclei with $A>2$ in Eq.~(\ref{14})  are expressed as follows \cite{wang2021nucleon}
\begin{equation}\label{15}
	n_{\text {corr }}(k) \cong \tau C_A\left[n_2(k)+n_5(k)\right],
\end{equation}
where $\tau$ is the proton number $Z$ or neutron number $N$. The parameter $C_A$ in Eq.~(\ref{15}) is the scaling factor of correlation strength  representing the ratio of high-momentum components between deuteron and other nuclei. $n_2(k)$ is derived from the second LFD wave function, which is primarily introduced by the tensor force and dominates in the region $1.5<k<3~\mathrm{fm^{-1}}$. $n_5(k)$ is obtained from the fifth LFD wave function, which is mainly induced by the $\pi$ exchange and dominates in $k>2.5~\mathrm{fm^{-1}}$.
%%%%%%%%%%%%%%%%%%%%%%%%%%%%%%%%%%%%%%%%%%%%%%%%%%%%%%%%%%%%%%%	
%%%%%%%%%%%%%%%%%%%%%%%%%%%%%%%%%%%%%%%%%%%%%%%%%%%%%%%%%%%%%%%	
\section{Numerical results and discussions}\label{sec:3}
%%%%%%%%%%%%%%%%%%%%%%%%%%%%%%%%%%%%%%%%%%%%%%%%%%%%%%%%%%%
%%%%%%%%%%%%%%%%%%%%%%%%%%%%%%%%%%%%%%%%%%%%%%%%%%%%%%%%%%%

In this section, we utilize the $\psi^{\prime}$ scaling function with the momentum distributions $n(k)$ to investigate the inclusive electron scattering $(\textit{e}, \textit{e}^{\prime})$. Moreover, we adopt the improved $\psi^{\prime}$ scaling function to analyze the influences of high-momentum nucleons on $(\textit{e}, \textit{e}^{\prime})$ cross sections.

%%%%%%%%%%%%%%%%%%%%%%%%%%%%%%%%%%%%%%%%%%%%%%%%%%%%%%%%%%%
%%%%%%%%%%%%%%%%%%%%%%%%%%%%%%%%%%%%%%%%%%%%%%%%%%%%%%%%%%%
\subsection{Scaling analyses on $(\textit{e}, \textit{e}^{\prime})$ in $r$- and $k$- spaces}	
%%%%%%%%%%%%%%%%%%%%%%%%%%%%%%%%%%%%%%%%%%%%%%%%%%%%%%%%%%%
%%%%%%%%%%%%%%%%%%%%%%%%%%%%%%%%%%%%%%%%%%%%%%%%%%%%%%%%%%%

\begin{figure}[H]
	\centering
	\includegraphics[width=0.5\textwidth]{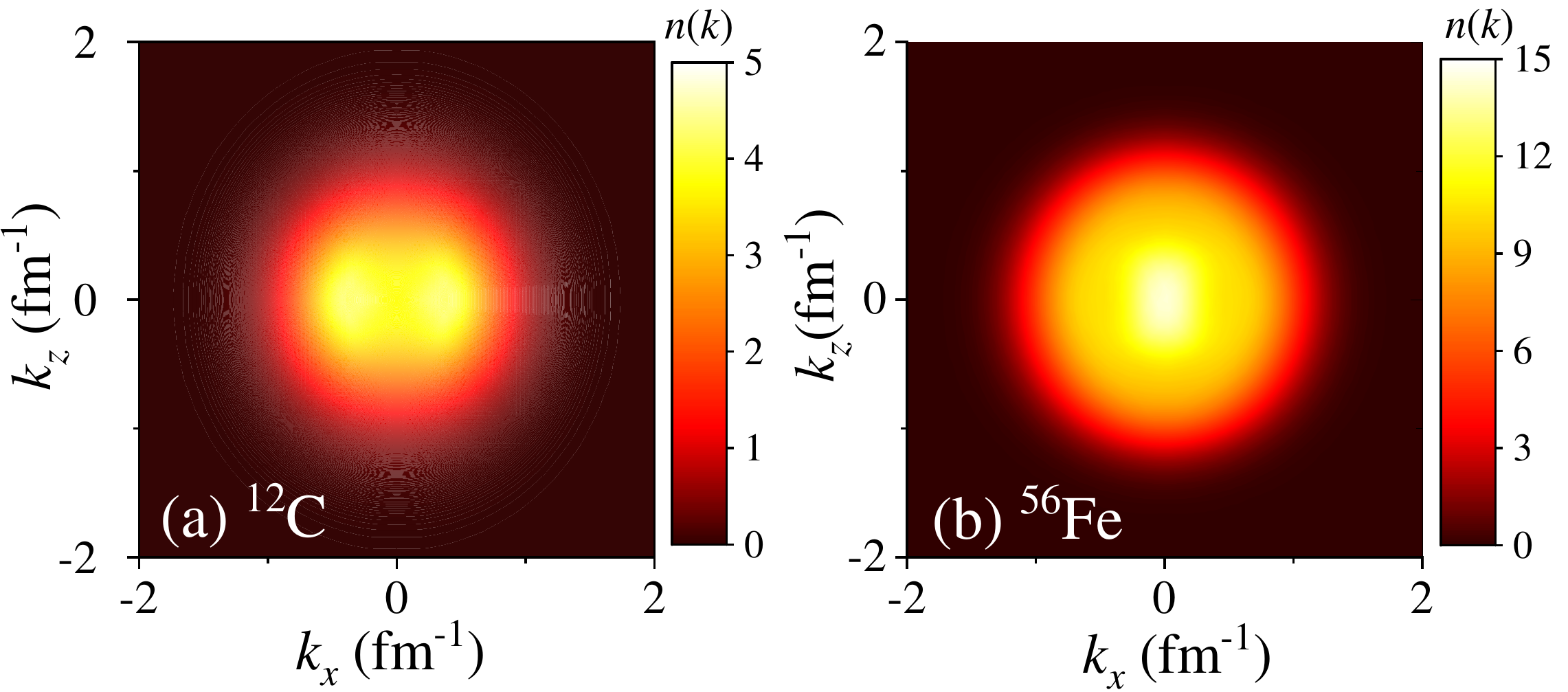}
	\caption{(a) Mean-field NMDs $n_{\mathrm{MF}}(k)$ $(\rm{fm}^3)$ of $^{12}\rm{C}$ for the nuclear deformation $\beta_{2}$~=~-0.33 calculated from the deformed HFB model with SLY4 parameters. (b) Same as (a) but for $^{56}\rm{Fe}$ with $\beta_{2}$~=~0.17.  The NMDs satisfy the normalization $\int n(k)~{d}^3k=A$. }
	\label{fig1}
\end{figure}

In Fig.~\ref{fig1}, the mean-field parts of nucleon momentum distributions (NMDs) $n_{\mathrm{MF}}(k)$ of $^{12}\rm{C}$ and $^{56}\rm{Fe}$ with different configurations are presented, which are calculated from the axially deformed HFB model. The Skyrme parameter set SLY4 is chosen in this paper \cite{chabanat1998skyrme}. Through the HFB calculations,  the deformation parameter of $^{12}\rm{C}$ is $\beta_{2}=-0.33$.  In Fig.~\ref{fig1}(a), the mean-field NMDs $n_{\mathrm{MF}}(k)$ of $^{12}\rm{C}$ show a localization pattern, which is due to the nuclear deformation effects. It is noted that the deformation can increase the nuclear potential depth and makes the wave functions more localized \cite{ebran2012atomic}, which is favorable for the formation of localization. In previous studies, the localization phenomenon in deformed mean-field calculations has been discussed systematically in $r$-space \cite{ebran2012atomic,Ebran2014Cluster,Sun2022psg}. One can observe from Fig.~\ref{fig1}(a) that the deformation also results in the localization in $k$-space. As shown in Fig.~\ref{fig1}(b), the mean-field NMDs $n_{\mathrm{MF}}(k)$ of $^{56}\rm{Fe}$ also exhibit a prolate shape for the deformation parameter $\beta_{2}$ = 0.17, which is in accordance with the experimental data \cite{raman2001transition}. 

\begin{figure*}[t]%t b 
	\centering
	\includegraphics[width=0.80\textwidth]{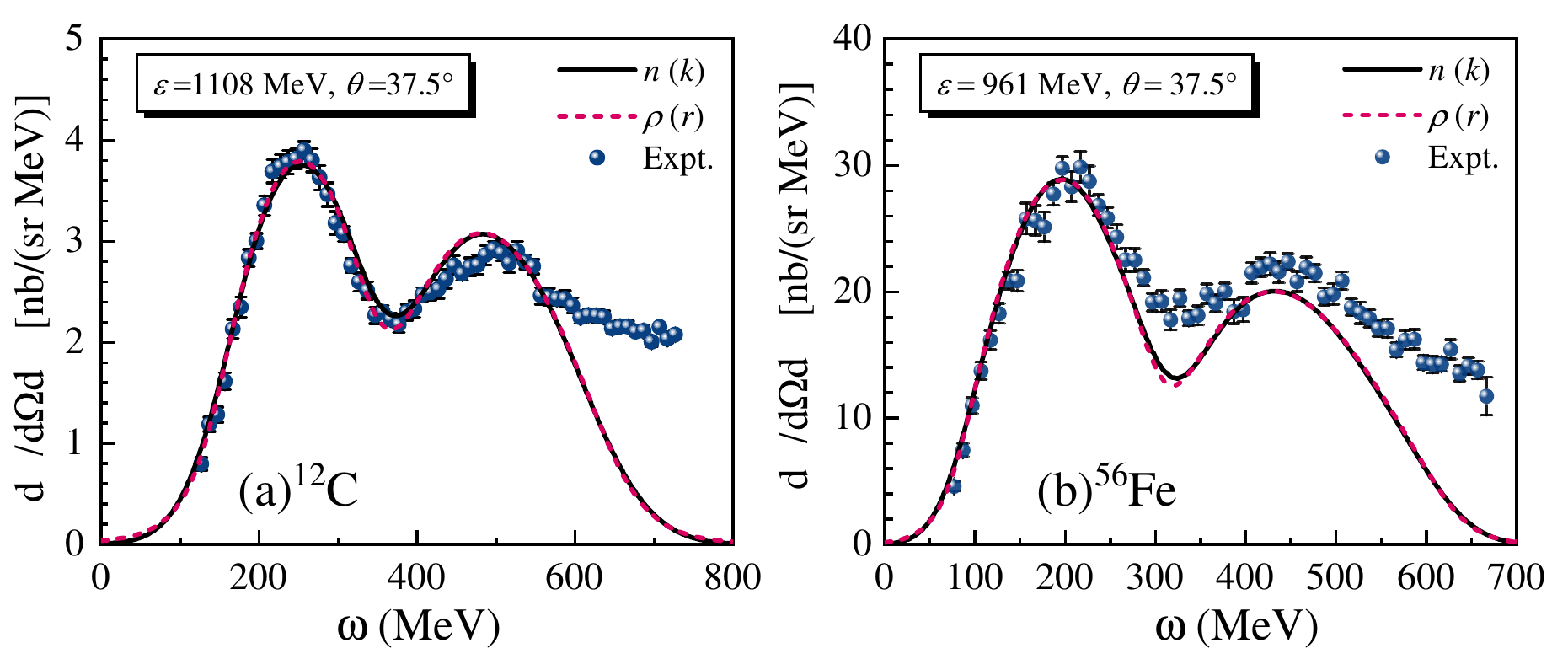}
	\caption{ (a) The inclusive electron scattering cross sections of $^{12}\rm{C}$ at the incident energy $\varepsilon$~=~1108 MeV and the scattering angle $\theta=37.5^{\circ}$. The results are calculated by two CDFM scaling functions $f_{\rho(r)}^{\rm{Q E}}\left(\psi^{\prime}\right)$ (dashed lines) and $f_{n(k)}^{\rm{Q E}}\left(\psi^{\prime}\right)$ (solid lines), where the corresponding $\rho(r)$ and $n(k)$ are from the deformed HFB calculations. (b) Same as (a) but for $^{56}\rm{Fe}$ at the kinematic $\varepsilon= 961$ MeV and $\theta=37.5^{\circ}$. The experimental data (balls) are taken from Ref. \cite{Sealock1989Electro}. }
	\label{fig2}
\end{figure*}
Based on NMDs in Fig.~\ref{fig1}, we further investigate the ($\textit{e},\textit{e}^{\prime}$) cross sections of $^{12}\rm{C}$ and $^{56}\rm{Fe}$ by utilizing the CDFM scaling function in $k$-space $f_{n(k)}^{\rm{Q E}}\left(\psi^{\prime}\right)$. In our calculations, we replace the parameter 0.75 to 0.72 in Eq.~(\ref{3}) for the QE region and change the parameter 0.75 to 0.54  in Eq.~(\ref{6}) for the $\Delta$ region, which can provide reasonable descriptions for the experimental data. The effects of Pauli blocking are neglected in this paper, as the kinematical setup corresponds to momentum transfer $q>500$ MeV/c \cite{BENHAR2007Total}. In Fig.~\ref{fig2}(a), the ($\textit{e},\textit{e}^{\prime}$) cross sections of $^{12}\rm{C}$ are presented at the incident energy $\varepsilon= 1108$ MeV and the scattering angle $\theta=37.5^{\circ}$. 
The theoretical results from the $\psi^{\prime}$ scaling function in $r$-space $f_{\rho(r)}^{\rm{QE}}\left(\psi^{\prime}\right)$ are also presented in Fig.~\ref{fig2}(a), where $\rho(r)$ is obtained from the HFB model with SLY4 parameters. As can be seen from Fig.~\ref{fig2}(a), the agreement between theoretical results and the experimental data is satisfactory, allowing us to verify the reliability of scaling function  $f_{n(k)}^{\rm{Q E}}\left(\psi^{\prime}\right)$ in $k$-space. Further comparing results of the scaling functions $f_{n(k)}^{\rm{Q E}}\left(\psi^{\prime}\right)$ and $f_{\rho(r)}^{\rm{Q E}}\left(\psi^{\prime}\right)$, one can see that two theoretical calculations are indistinguishable, which demonstrates that two scaling functions in different spaces are equivalent. 

The ($\textit{e},\textit{e}^{\prime}$) cross sections of $^{56}\rm{Fe}$ at the kinematic $\varepsilon~=~961$~MeV and $\theta=37.5^{\circ}$ are also  provided in Fig.~\ref{fig2}(b). Similar to Fig.~\ref{fig2}(a), the consistency between $f_{n(k)}^{\rm{Q E}}\left(\psi^{\prime}\right)$ and $f_{\rho(r)}^{\rm{Q E}}\left(\psi^{\prime}\right)$ is clearly illustrated in Fig.~\ref{fig2}(b). There are two reasons to explain these phenomena. On the one hand, the momentum distribution $n(k)$ can be translated from density distribution $\rho(r)$ by the self-consistent Fourier transform of the nucleon wave functions from $r$-space $\Phi_\alpha(\boldsymbol{r} , \sigma)$ to $k$-space $\tilde\Phi_\alpha(\boldsymbol{k} , \sigma)$. On the other hand, within the framework of CDFM, $f_{\rho(r)}^{\rm{Q E}}\left(\psi^{\prime}\right)$ of Eq.~(45) in Ref.~\cite{Antonov2005Superscaling} and $f_{n(k)}^{\rm{Q E}}\left(\psi^{\prime}\right)$ of Eq.~(\ref{2}) in this paper can be turned into each other by the integral-variable transformation from $r$-space to $k$-space. Therefore, the scaling functions $f_{n(k)}^{\rm{Q E}}\left(\psi^{\prime}\right)$ and $f_{\rho(r)}^{\rm{Q E}}\left(\psi^{\prime}\right)$ are essentially the same physical quantities in different spaces.

%%%%%%%%%%%%%%%%%%%%%%%%%%%%%%%%%%%%%%%%%%%%%%%%%%%%%%%%%%%
%%%%%%%%%%%%%%%%%%%%%%%%%%%%%%%%%%%%%%%%%%%%%%%%%%%%%%%%%%%
\subsection{$(\textit{e},\textit{e}^{\prime}$) cross section from improved $\psi^{\prime}$ scaling function }
%%%%%%%%%%%%%%%%%%%%%%%%%%%%%%%%%%%%%%%%%%%%%%%%%%%%%%%%%%%
%%%%%%%%%%%%%%%%%%%%%%%%%%%%%%%%%%%%%%%%%%%%%%%%%%%%%%%%%%%

\begin{figure}[H] %t b 
	\centering
	\includegraphics[width=0.43\textwidth]{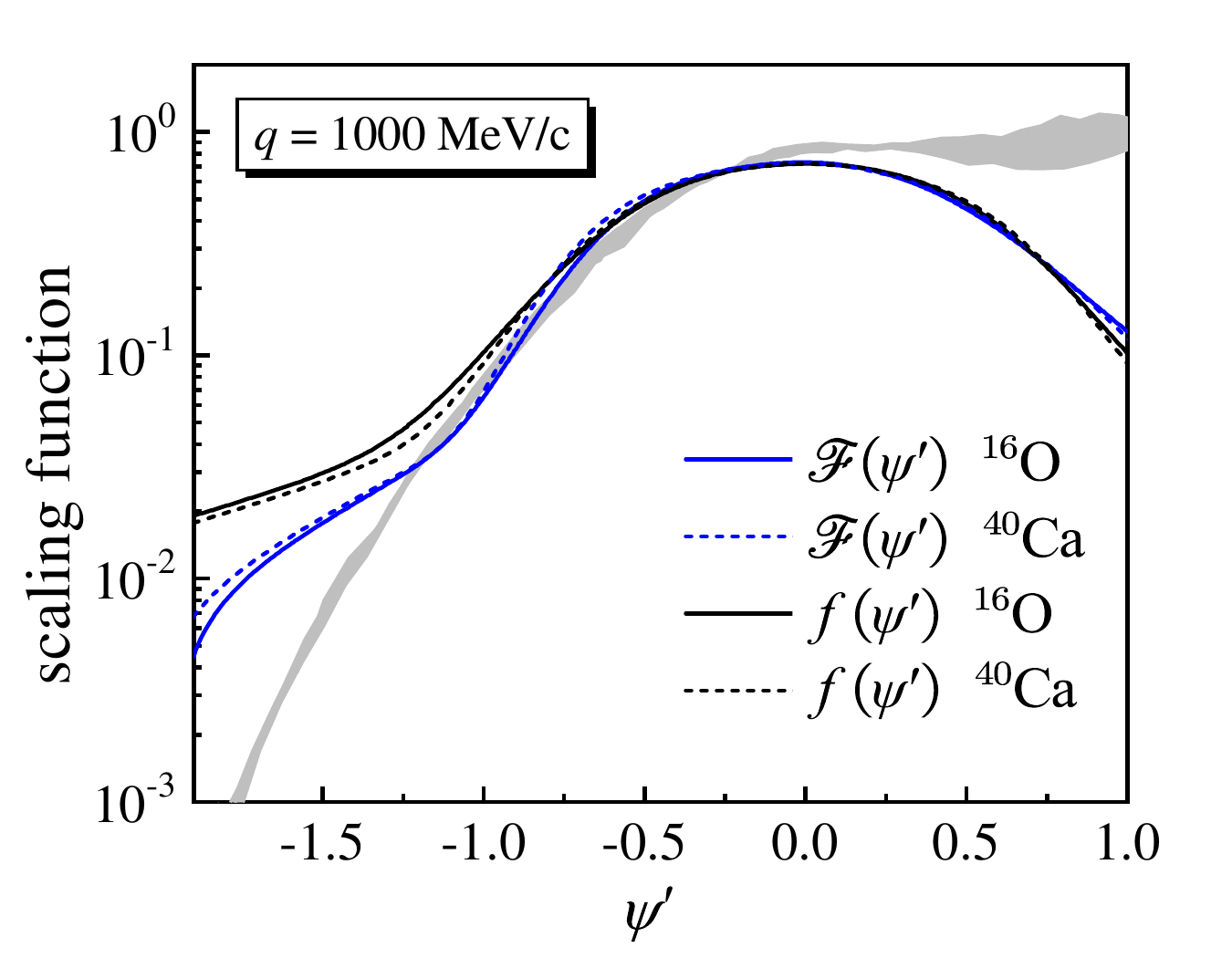}
	\caption{ The $\psi^{\prime}$ scaling function $f_{n(k)}^{\rm{Q E}}\left(\psi^{\prime}\right)$ (black lines) and improved $\psi^{\prime}$ scaling function $\mathscr{F}_{n(k)}^{\rm{QE}}\left(\psi^{\prime}\right)$ (blue lines) in $k$-space at $q$~=~1000 MeV/c for $^{16}\rm{O}$ and $^{40}\rm{Ca}$, where $n(k)$ is obtained from HFB+LFD calculations. The experimental data (gray area) are from Ref. \cite{Donnelly1999Superscaling}.  }
	\label{fig5}
\end{figure}

In this part, the improved scaling function in $k$-space $\mathscr{F}_{n(k)}^{\rm{QE}}\left(\psi^{\prime}\right)$ is applied to investigate the ($\textit{e},\textit{e}^{\prime}$) cross sections, where the corresponding nucleon momentum distributions $n(k)$ are corrected by introducing the \textit{NN}-SRC effects. The mean-field part $n_{\mathrm{MF}}(k)$ is from the HFB calculations with Eq.~(\ref{13}), and the high-momentum component  $n_{\mathrm{corr}}(k)$ of NMD is included by the LFD method with Eq.~(\ref{15}).  In Fig.~\ref{fig5}, we provide two scaling functions $f_{n(k)}^{\rm{Q E}}\left(\psi^{\prime}\right)$ and $\mathscr{F}_{n(k)}^{\rm{QE}}\left(\psi^{\prime}\right)$ for $^{16}\rm{O}$ and $^{40}\rm{Ca}$ at momentum transfer $q$~=~1000 MeV/c. As shown in this figure, both two scaling functions coincide with the experimental data in the whole trend. Considering the influences of high-momentum nucleons, there are different behaviors in the negative-$\psi^{\prime}$ region that corresponds to the low-$\omega$ region of the ($\textit{e},\textit{e}^{\prime}$) cross sections. The values of $\mathscr{F}_{n(k)}^{\rm{QE}}\left(\psi^{\prime}\right)$ are smaller than $f_{n(k)}^{\rm{Q E}}\left(\psi^{\prime}\right)$ , especially in the region  $\psi^{\prime}<-1.0$ of Fig.~\ref{fig5}.

\begin{figure*}[htbp] %t b 
	\centering
	\includegraphics[width=0.80\textwidth]{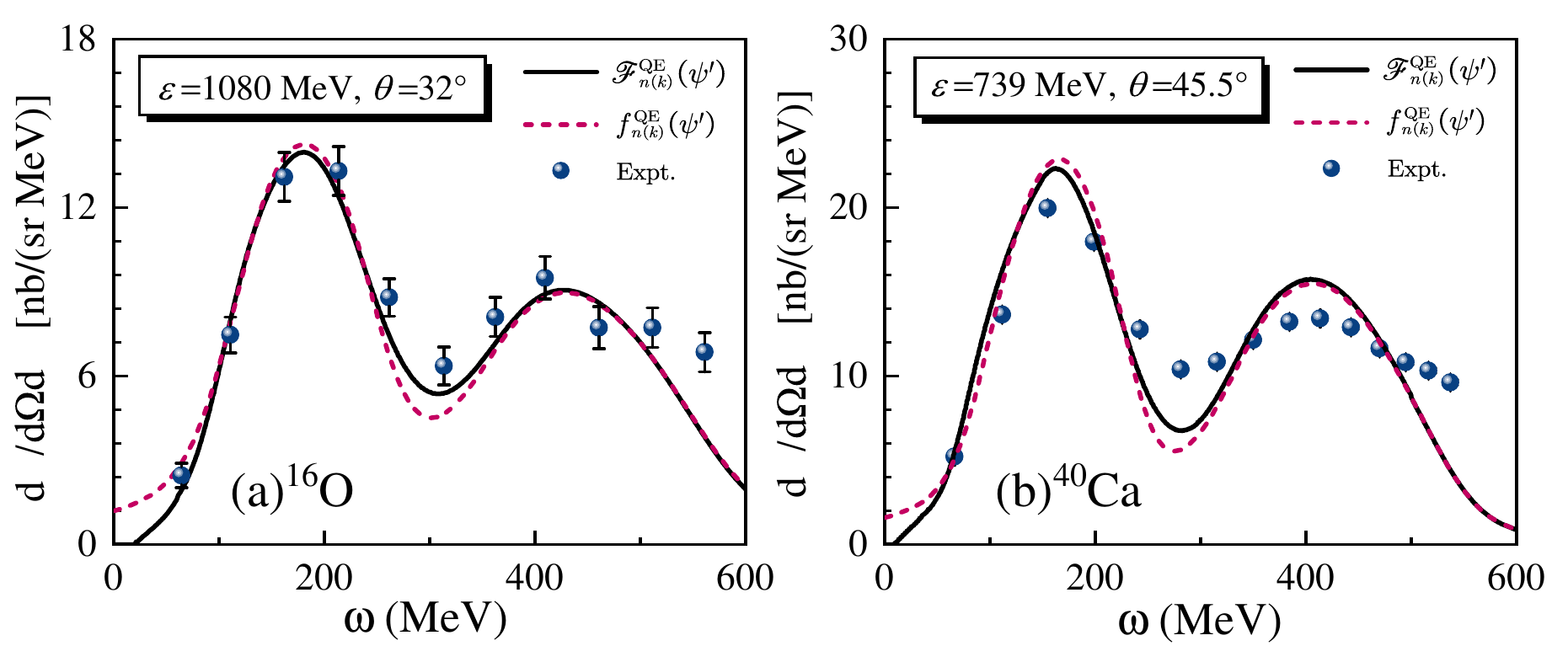}
	\caption{ (a) The inclusive electron scattering cross sections of $^{16}\rm{O}$ at the incident energy $\varepsilon$~=~1080~MeV and the scattering angle $\theta$~=~$32^{\circ}$, calculated by $\mathscr{F}_{n(k)}^{\rm{QE}}\left(\psi^{\prime}\right)$ (solid lines) and $f_{n(k)}^{\rm{Q E}}\left(\psi^{\prime}\right)$ (dashed lines) respectively, where $n(k)$ is obtained from HFB+LFD calculations. (b) Same as (a) but for $^{40}\rm{Ca}$ at the kinematic $\varepsilon=739$ MeV and $\theta=45.5^{\circ}$. The experimental data (balls) are from Refs. \cite{Anghinolfi1996Quasi,Williamson1997Quasielastic}. }
	\label{fig3}
\end{figure*}

In Fig.~\ref{fig3}, we provide the ($\textit{e},\textit{e}^{\prime}$) cross sections calculated by two scaling functions  $f_{n(k)}^{\rm{Q E}}\left(\psi^{\prime}\right)$ and $\mathscr{F}_{n(k)}^{\rm{QE}}\left(\psi^{\prime}\right)$ for $^{16}\rm{O}$ and $^{40}\rm{Ca}$, where $n(k)$ is from the HFB+LFD calculations. As shown in this figure, the results obtained from $f_{n(k)}^{\rm{Q E}}\left(\psi^{\prime}\right)$ and $\mathscr{F}_{n(k)}^{\rm{QE}}\left(\psi^{\prime}\right)$ agree with the experimental data. Both of them pass through the smallest ($\textit{e},\textit{e}^{\prime}$) data. Due to the different integral bounds of scaling functions in Eqs. (\ref{2}) and (\ref{9}), the two scaling analyses lead to different cross sections in the region $\omega \simeq 0$ of Fig.~\ref{fig3}, when the high-momentum component of $n(k)$ is included.  The ($\textit{e},\textit{e}^{\prime}$) cross sections from $f_{n(k)}^{\rm{Q E}}\left(\psi^{\prime}\right)$ are greater than zero at the energy transfer $\omega=0$. However, for $\omega=0$, only the elastic electron scattering occurs, which means the ($\textit{e},\textit{e}^{\prime}$) cross sections should be zero. The inconsistency of cross sections in the low-$\omega$ region indicates that $f_{n(k)}^{\rm{Q E}}\left(\psi^{\prime}\right)$ needs to be improved.

%Results of the improved scaling function  $\mathscr{F}_{n(k)}^{\rm{QE}}\left(\psi^{\prime}\right)$ are presented in Fig.~\ref{fig3} for $^{16}\rm{O}$ and $^{40}\rm{Ca}$, where the $n(k)$ are from the HFB+LFD calculations. 

As shown in Fig.~\ref{fig3}, the ($\textit{e},\textit{e}^{\prime}$) cross sections from the improved scaling function in $k$-space $\mathscr{F}_{n(k)}^{\rm{QE}}\left(\psi^{\prime}\right)$  vanish at $\omega=0$, which satisfies the physical mechanism that only elastic scattering occurs. This can be interpreted from the limits of integration of $\mathscr{F}_{n(k)}^{\rm{QE}}\left(\psi^{\prime}\right)$ in Eq.~(\ref{9}). Further calculating the integral limits of Eq.~(\ref{9}), one can obtain that in the region $\omega<40$ MeV of Fig.~\ref{fig3}, only the nucleons with momentum $k>2 ~\text{fm}^{-1}$ can participate in the ($\textit{e},\textit{e}^{\prime}$) process. It indicates that the low-$\omega$ region of ($\textit{e},\textit{e}^{\prime}$) cross sections is more sensitive to the high-momentum nucleons within the framework of $\mathscr{F}_{n(k)}^{\rm{QE}}\left(\psi^{\prime}\right)$. Therefore, the improved scaling function $\mathscr{F}_{n(k)}^{\rm{QE}}\left(\psi^{\prime}\right)$ has the ability to analyze the influences of \textit{NN}-SRC on the ($\textit{e},\textit{e}^{\prime}$) process. 
%%%%%%%%%%%%%%%%%%%%%%%%%%%%%%%%%%%%%%%%%%%%%%%%%%%%%%%%%%%
%%%%%%%%%%%%%%%%%%%%%%%%%%%%%%%%%%%%%%%%%%%%%%%%%%%%%%%%%%%
\subsection{\textit{NN}-SRC effects on ($\textit{e},\textit{e}^{\prime}$) cross section}
%%%%%%%%%%%%%%%%%%%%%%%%%%%%%%%%%%%%%%%%%%%%%%%%%%%%%%%%%%%
%%%%%%%%%%%%%%%%%%%%%%%%%%%%%%%%%%%%%%%%%%%%%%%%%%%%%%%%%%%

\begin{figure*}[htb] %t b 
	\centering
	\includegraphics[width=0.9\textwidth]{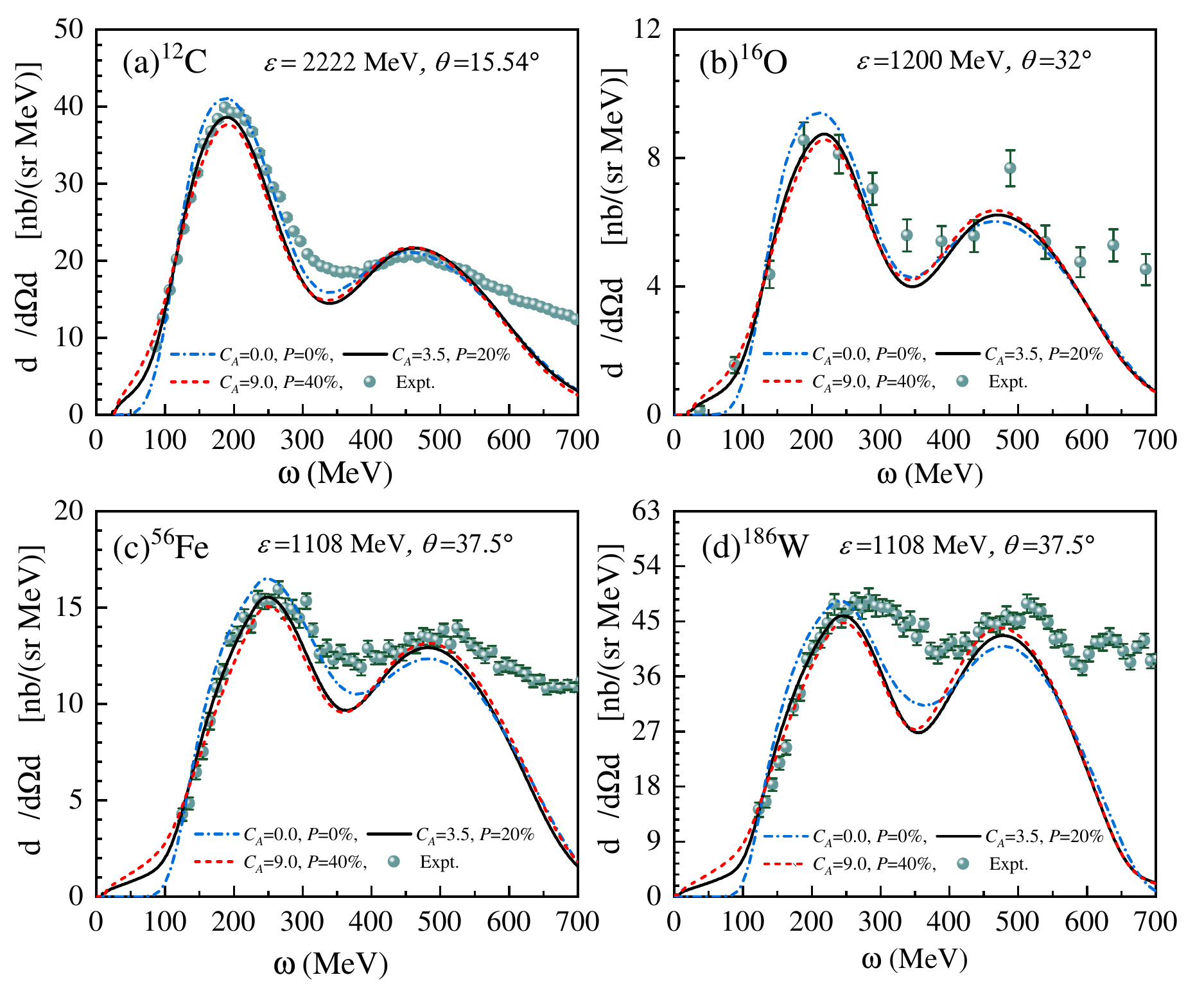}
	\caption{ (a) The inclusive cross sections of $^{12}\rm{C}$ at the kinematic $\varepsilon=2222$ MeV and $\theta=15.54^{\circ}$ for different $\textit{NN}$-SRC strengths calculated by  $\mathscr{F}_{n(k)}^{\rm{QE}}\left(\psi^{\prime}\right)$, where $n(k)$ is from the HFB+LFD method. (b) Same as panel (a) but for $^{16}\rm{O}$ at $\varepsilon=1200$ MeV and $\theta=32^{\circ}$. (c) Same as panel (a) but for $^{56}\rm{Fe}$ at $\varepsilon=1108$ MeV and $\theta=37.5^{\circ}$. (d) Same as panel (a) but for $^{186}\rm{W}$ at $\varepsilon=1108$ MeV and $\theta=37.5^{\circ}$. The experimental data are taken from Refs. \cite{Anghinolfi1996Quasi,Sealock1989Electro,Dai2019First} .}
	\label{fig4}
\end{figure*}

In this part, we apply the improved scaling function $\mathscr{F}_{n(k)}^{\rm{QE}}\left(\psi^{\prime}\right)$ to investigate the $\textit{NN}$-SRC effects on the inclusive electron scattering. The ($\textit{e},\textit{e}^{\prime}$) cross sections of selected nuclei $^{12}\rm{C}$, $^{16}\rm{O}$, $^{56}\rm{Fe}$, and $^{186}\rm{W}$ are calculated and presented in Fig.~\ref{fig4}, where the $\textit{NN}$-SRC effects are introduced by the LFD method. The parameters $C_{A}$ of Eq.~(\ref{15}) are chosen to obtain the different $\textit{NN}$-SRC strength.
%The different correlation strengths $C_{A}$ of Eq.~(\ref{15}) are chosen to obtain the different proportions of correlated nucleons. 
As a ratio of high-momentum parts $n_\text{corr}(k)$ to the total $n(k)$, the quantity $P$ in Fig.~\ref{fig4} represents the percentage of correlated nucleons in total nucleon number. One can see that in both QE and $\Delta$ resonance regions of Fig.~\ref{fig4}, the theoretical ($\textit{e},\textit{e}^{\prime}$) cross sections coincide with the experimental data in a satisfactory way. It should be mentioned that the results of $^{12}\rm{C}$ agree well with the latest high-precision data, which were measured in the E12-14-012 experiment at Jefferson
Laboratory \cite{Dai2018First,Dai2019First}. This indicates the improved scaling function in $k$-space $\mathscr{F}_{n(k)}^{\rm{QE}}\left(\psi^{\prime}\right)$ can successfully reproduce ($\textit{e},\textit{e}^{\prime}$) data in a variety of kinematic regimes for different nuclei. There are still certain discrepancies between the theoretical results and experimental data in the high-$\omega$ region. This is due to the contributions of deep inelastic electron scattering, where the electron interacts with quarks inside nucleons.

To analyze the $\textit{NN}$-SRC influences on ($\textit{e},\textit{e}^{\prime}$) cross sections, the correlation strengths are chosen to $C_{A}$~=~0, 3.5, and 9.0 for the selected nuclei, and the corresponding proportions of correlated nucleons are $P$~=~0\%, 20\%, and 40\%, respectively. As displayed in Fig.~\ref{fig4}, there are discrepancies in the ($\textit{e},\textit{e}^{\prime}$) cross sections for different $\textit{NN}$-SRC strengths. In the low-$\omega$ region, the ($\textit{e},\textit{e}^{\prime}$) cross sections increase with the \textit{NN}-SRC strength. This is due to the fact that the high-momentum nucleons primarily contribute  to the low-$\omega$ region of the ($\textit{e},\textit{e}^{\prime}$) cross section.
Additionally, it is straightforward to see that the increase of $\textit{NN}$-SRC strength causes a reduction of the theoretical ($\textit{e},\textit{e}^{\prime}$) cross sections around the QE peak, which can be attributed to the integral limits of $k$ in  Eq.~(\ref{9}). As seen in Fig.~\ref{fig4}, the improved $\psi^{\prime}$ scaling function  $\mathscr{F}_{n(k)}^{\rm{QE}}\left(\psi^{\prime}\right)$ is appropriate for analyzing the sensitivity of \textit{NN}-SRC strength on ($\textit{e},\textit{e}^{\prime}$) cross sections. 
% From Fig.~\ref{fig4}, one can see that the improved scaling function $\mathscr{F}_{n(k)}^{\rm{QE}}\left(\psi^{\prime}\right)$ is suitable to investigate the sensitivity of $\textit{NN}$-SRC strength on ($\textit{e},\textit{e}^{\prime}$) cross sections.

%The ($\textit{e},\textit{e}^{\prime}$) cross sections with various correlation strengths $C_A$ in Fig.~\ref{fig4} are further analyzed for the selected nuclei. 

%For the purpose of extracting the correlation strengths $C_A$, the ($\textit{e},\textit{e}^{\prime}$) cross sections calculated by the improved scaling function $\mathscr{F}_{n(k)}^{\rm{QE}}\left(\psi^{\prime}\right)$ with various $C_A$ are further analyzed in Fig.~\ref{fig4}.  
On the basis of the improved scaling function $\mathscr{F}_{n(k)}^{\rm{QE}}\left(\psi^{\prime}\right)$, the proportions of correlated nucleons in nuclei are further analyzed from the ($\textit{e},\textit{e}^{\prime}$) cross sections. As can be seen in Fig.~\ref{fig4}(a) that $\mathscr{F}_{n(k)}^{\rm{QE}}\left(\psi^{\prime}\right)$ with $C_A=3.5$ provides reasonable ($\textit{e},\textit{e}^{\prime}$) cross sections for $^{12}\rm{C}$, which coincides with the experimental data. The QE peak is overestimated by $C_A=0.0$  whereas it is underestimated by the calculations of $C_A=9.0$. By comparing  the ($\textit{e},\textit{e}^{\prime}$) cross sections in Fig.~\ref{fig4}(a), one can extract the proportion of correlated nucleons $P$~=~20\% for $^{12}\rm{C}$. This result is supported by the latest exclusive electron scattering experiment that 20\% of the nucleons move in the form of nucleon-nucleon pairs \cite{Subedi2008Probing,Hen2014Momentum}. Previous \textit{ab initio} calculations also indicated that the correlated nucleons account for approximately 20\% contributions of the total nucleons \cite{Hen2017Nucleon,LYU2020high}. For the results of $^{16}\rm{O}$, $^{56}\rm{Fe}$, and $^{186}\rm{W}$ in Fig.~\ref{fig4}, the theoretical ($\textit{e},\textit{e}^{\prime}$) cross sections with $P$~=~20\% are all in accord with the experimental data. Through the generalized analyses of the results of Fig.~\ref{fig4}, one can observe that the improved scaling function $\mathscr{F}_{n(k)}^{\rm{QE}}\left(\psi^{\prime}\right)$ can be used to investigate the $\textit{NN}$-SRC effects in nuclei.

%%%%%%%%%%%%%%%%%%%%%%%%%%%%%%%%%%%%%%%%%%%%%%%%%%%%%%%%%%%%%%%%%%%%%%%
%%%%%%%%%%%%%%%%%%%%%%%%%%%%%%%%%%%%%%%%%%%%%%%%%%%%%%%%%%%%%%%%%%%%%%%
\section{Conclusion}\label{sec:4}

Quasielastic electron scattering has played an essential role in the research of nuclear structure. As a powerful tool, scaling analysis is widely used in investigating the  inclusive electron scattering ($\textit{e},\textit{e}^{\prime}$) cross sections. In this paper, we propose an improved scaling function $\mathscr{F}_{n(k)}^{\rm{QE}}\left(\psi^{\prime}\right)$ in $k$-space  to investigate the nucleon momentum distributions $n(k)$ and analyze the $\textit{NN}$-SRC effects. 

Firstly, the  scaling functions in $r$-space $f_{\rho(r)}^{\rm{QE}}\left(\psi^{\prime}\right)$ and $k$-space $f_{n(k)}^{\rm{QE}}\left(\psi^{\prime}\right)$ are compared with each other, where the density distribution $\rho(r)$ and momentum distribution $n(k)$ are from the axially deformed HFB model. The consistency between $f_{n(k)}^{\rm{QE}}\left(\psi^{\prime}\right)$ and $f_{\rho(r)}^{\rm{QE}}\left(\psi^{\prime}\right)$ is clearly illustrated, which indicates $f_{n(k)}^{\rm{QE}}\left(\psi^{\prime}\right)$ and $f_{\rho(r)}^{\rm{QE}}\left(\psi^{\prime}\right)$ are the same physical quantities in different spaces. Next, considering the momentum and energy conservation in the ($\textit{e},\textit{e}^{\prime}$) process, an improved scaling function $\mathscr{F}_{n(k)}^{\rm{QE}}\left(\psi^{\prime}\right)$ is constructed in $k$-space to study the asymptotic behavior of $n(k)$ at the high-$k$ region. The comparison between $f_{n(k)}^{\rm{QE}}\left(\psi^{\prime}\right)$ and $\mathscr{F}_{n(k)}^{\rm{QE}}\left(\psi^{\prime}\right)$ indicates that $\mathscr{F}_{n(k)}^{\rm{QE}}\left(\psi^{\prime}\right)$ is sensitive to the high-momentum component of $n(k)$, which has ability to investigate the $\textit{NN}$-SRC effects. Finally, on the basis of $\mathscr{F}_{n(k)}^{\rm{QE}}\left(\psi^{\prime}\right)$, the strengths of $\textit{NN}$-SRC for the selected nuclei are extracted from the ($\textit{e},\textit{e}^{\prime}$) cross sections. This paper proposes a new approach to investigate the inclusive electron scattering, which is also helpful for the studies of neutrino-nucleus scattering.

\Acknowledgements{The authors are grateful to A. N. Antonov and M. V. Ivanov for valuable discussions and careful reading of the manuscript.  This work is supported by the National Natural Science Foundation of China (Grants No. 12035011, No. 11535004, No. 11975167, No. 11761161001, No. 11565010, No. 11961141003, and No. 12022517), by the National Key R$\&$D Program of China (Contracts No. 2018YFA0404403 and No. 2016YFE0129300), by the Science and Technology Development Fund of Macau (Grants No. 0048/2020/A1 and No. 008/2017/AFJ), by the Shan dong Provincial Natural Science Foundation, China (Grant No. ZR2020MA096), by the Open Project of Guangxi Key Laboratory of Nuclear Physics and Nuclear Technology (Grant No. NLK2021-03), by the Key Laboratory of High Precision Nuclear Spectroscopy, Institute of Modern Physics, Chinese Academy of Sciences (Grant No. IMPK FKT2021001), and by the Fundamental Research Funds for the Central Universities (Grant No. 22120200101).}

\bibliography{references}
\end{multicols}
\end{document}